\documentclass[11pt,a4paper]{article}
\pdfoutput=1
\usepackage{jheppub}

\usepackage{epsf,epsfig,amsmath,amssymb,dsfont}
\usepackage{amsthm,mathrsfs} 
\usepackage{graphicx} 
\usepackage{multirow}
\usepackage{float}
\usepackage{colortbl}
\usepackage[table]{xcolor}
\usepackage{varioref}

\newcommand{\be}{\begin{equation}}
\newcommand{\ee}{\end{equation}}
\newcommand{\baln}{\begin{align}}
\newcommand{\ealn}{\end{align}}
\newcommand{\ben}{\begin{equation*}}
\newcommand{\een}{\end{equation*}}

\long\def\symbolfootnote[#1]#2{\begingroup%
\def\thefootnote{\fnsymbol{footnote}}\footnote[#1]{#2}\endgroup}

\newcommand{\nn}{\nonumber \\}

\newcommand{\fr}{\frac}
\newcommand{\del}{\partial}
\newcommand\CS{\mathcal{C}}

\newcommand{\vac}{|0\rangle}
\newcommand{\dg}{\delta^+G(k)}
\newcommand{\suppdg}{\text{supp}(\delta^+G(k))}
\newcommand{\kvec}{\bold{k}}
\newcommand{\aomega}{a(\Omega,\kvec)}

\newcommand{\dk}{d^{3}\textbf{k}}

\newcommand{\dy}{d^{3}\textbf{y}}
\newcommand{\dz}{d^{3}\textbf{z}}
\newcommand{\dzp}{d^{3}\textbf{z}'}
\newcommand{\Dy}{\frac{\partial}{\partial y^{0}}}
\newcommand{\Dz}{\frac{\partial}{\partial z^{0}}}
\newcommand{\Dzp}{\frac{\partial}{\partial z'^{0}}}
\newcommand{\Dm}{\Delta_{m^2}}
\newcommand{\Dmz}{\Delta_{m_{0}^2}}
\newcommand{\Dmu}{\Delta_{\mu^2}}

\title{Nonlocal Scalar Quantum Field Theory from Causal Sets}

\author[a,b]{Alessio Belenchia,} 
\author[a,b]{Dionigi M. T. Benincasa}
\author[a,b]{and Stefano Liberati}

\affiliation[a]{SISSA - International School for Advanced Studies, Via Bonomea 265, 34136 Trieste, Italy.}
\affiliation[b]{INFN, Sezione di Trieste, Via Valerio 2, Trieste, 34127 Italy.}

\emailAdd{dionigi.benincasa@sissa.it}
\emailAdd{abelen@sissa.it}
\emailAdd{liberati@sissa.it}

\abstract{
We study a non-local scalar quantum field theory in flat spacetime derived from the 
dynamics of a scalar field on a causal set.
We show that this non-local QFT contains a continuum of massive modes in any dimension. 
In 2 dimensions the Hamiltonian is positive definite and therefore the quantum theory is well-defined.
In 4-dimensions, we show that the unstable modes of the non-local d'Alembertian are propagated via the so called Wheeler propagator and hence do not appear in the asymptotic states. In the free case studied here the continuum of massive mode are shown to not propagate in the asymptotic states.
However the Hamiltonian is not positive definite, therefore potential issues with the quantum theory remain. 
Finally, we conclude with hints toward what kind of phenomenology one
might expect from such non-local QFTs.
}

\begin{document}
\maketitle
\flushbottom


\section{Introduction}

Understanding the mesoscopic behaviour of quantum gravity (QG) theories 
is key to our ability to test them observationally.
Recent years have witnessed growing activity in the field of quantum gravity phenomenology 
\cite{liberati2013tests,Mattingly:2005kq,amelino2013quantum,hossenfelder2013minimal}, with 
unforeseeable success having been achieved in testing possible breakdowns of spacetime 
symmetries close to the Planck scale. In particular, tests of effective field theories (EFT) with 
ultraviolet (UV) departures from Local Lorentz Invariance have been proven invaluable
in providing severe constraints on such scenarios  \cite{liberati2013tests}. 
Nonetheless, the very same successes of these studies, together with observational advancements in 
cosmology, high energy astrophysics and particle physics have emphasised the need to link concrete 
QG proposals to observations. Unfortunately, many of the current QG models lack the theoretical
maturity necessary for this link to be established with certainty, although activity in this direction
has been growing lately. A few examples of such attempts are the emergent spacetime models within the 
Group Field Theory formalism of quantum gravity~\cite{Gielen:2014uga}, 
loop quantum gravity applications to cosmology~\cite{Ashtekar:2011ni} and black holes~\cite{Barrau:2014vn}, 
and asymptotic safety scenarios~\cite{lrr-2006-5}.  

Alongside these approaches, there have recently been interesting developments 
in Causal Set Theory (CST).
\footnote{For nice reviews of causal set theory with 
emphasis on modern developments see \cite{Surya:2011sf} and \cite{Henson:2006kf}.}
CST postulates that the fundamental structure of 
spacetime is a discrete/locally-finite partial order, whose order relation is taken to underlie 
the macroscopic causal
order of spacetime events. In order to marry fundamental discreteness with (local) Lorentz invariance the
causal set is subject to a kinematic randomness, made concrete
by the concept of {\it sprinkling}, i.e.  
a way of generating a causet from a 
$d$-dimensional Lorentzian manifold $(\mathcal{M},g)$. Sprinkling
is a Poisson process of selecting points in $\mathcal{M}$
with density $\rho=1/l^d$ ($l$ being the discreteness scale) 
so that the  expected number of points 
sprinkled in a region of spacetime volume $V$ is $\rho V$.
This process generates 
a causet whose elements are the sprinkled points and whose 
order is that induced by the manifold's causal 
order restricted to the sprinkled points. 
It is then said that a causet $\CS$ 
is well approximated by a manifold $(\mathcal{M},g)$ if
it could have been generated, with relatively 
high probability, by sprinkling into $(\mathcal{M},g)$.
 
A salient feature of the interplay between discreteness and Lorentz invariance is 
that locality has to be given up.
To understand why, consider the nearest neighbours to a given point, $p$, in a causal set 
well-approximated by Minkowski spacetime. 
These will lie roughly on the hyperboloid lying one Planck unit of proper time away from $p$ 
and therefore will be infinite in number. In the case where curvature
limits Lorentz symmetry this number may not be infinite but will still be huge.  
This inherent non-locality is evident in the definition of d'Alembert's operator on a causal set,
defined by constructing a finite difference equation
in which linear combinations of the value of the field at neighbouring points are taken. Since
the number of nearest neighbours, next nearest neighbours, etc., is very large (infinite), the
corresponding expression looks highly non-local (see for example Equation (2) of \cite{Benincasa:2010ac}). 
Nonetheless, the resulting operator
can be shown to be approximately local, with the non-locality confined to scales of order the
discreteness scale.
The precise form of this correspondence is given by performing an average of the causal
set d'Alembertian over all sprinklings
of a given spacetime~\footnote{In this paper we are only going to be interested in flat spacetimes, 
although similar results have been shown to hold in curved spacetimes \cite{Benincasa:2010ac}.}, 
giving rise to a non-local, retarded, Lorentz invariant linear operator in the continuum, $\Box_\rho$,
whose non-locality is parametrised by the discreteness scale $l$, and is such that it reduces
to the local continuum d'Alembertian, $\Box$, in the limit $l\rightarrow0$. Hence, $\Box_{\rho}$
encodes an averaged effect of the underlying spacetime discreteness on the propagation of
scalar fields.

It was soon realised however \cite{Sorkin:2007qi}, 
that although the mean of this operator has the correct continuum limit, 
its discrete counterpart suffers from large/unacceptable fluctuations (growing with $\rho$). 
This was solved by introducing a new length scale, $l_k\gg l$, 
over which the original discrete expressions were ``smeared out". When averaged over sprinklings
these operators lead to non-local d'Alembertians where the non-locality is now confined to scales
of order $l_k$ rather than $l$. As such, nonlocal corrections to the exact continuum d'Alembertian
are expected to still be relevant at scales where the continuum description of the spacetime 
is already valid to a good approximation. Again, locality is restored in the limit $l_k\rightarrow0$.
Note that at this stage the length scale $l_k$ enters as a purely phenomenological parameter, introduced
in order to make the discrete propagation physically meaningful. As such, $l_k$ plays a similar
role to length scales often introduced in quantum gravity phenomenology (see for example 
\cite{liberati2013tests} and references therein).

The focus of this paper is to study the non-local effective field theory for a scalar field whose
dynamics are given by the non-local d'Alembertian for which we take $l_k\ne0$. 
The mesoscopic regime where non-local effects start to play an important role are those of major interest 
and require a proper phenomenological study which we aim to address in a 
future publication.

The paper is organised as follows. In Section \ref{NLBOX} we review properties 
of the continuum non-local d'Alembertian derived from its CST counterpart.
In Section \ref{massive} we extend the definition of the non-local d'Alembertian to include
massive fields. Section \ref{huygens} is dedicated to the study of the retarded Green functions
found in \cite{Aslanbeigi:2014tg} in 2 and 4 dimensions and Huygens' principle.  
In Section \ref{FSNLQFT} we construct the effective field theory for a scalar field based on the 
aforementioned non-local d'Alembertian. In particular we will discuss the structure of the solutions 
in two and four dimensions. Finally, in Section \ref{summary} we discuss the physical 
implications of our study and suggest future perspectives about the phenomenology apt to constraint them.

\section{Nonlocal d'Alembertians}
\label{NLBOX}

The first construction of a d'Alembertian operator on a causal set 
appeared in a seminal paper by Sorkin \cite{Sorkin:2007qi}.
This was later extended to 4 dimensions and curved spacetimes,
\cite{Benincasa:2010ac}, and subsequently to all other dimensions 
\cite{Glaser:2013sf,Dowker:2013vl} and with an arbitrary
number of layers \cite{Aslanbeigi:2014tg}. 
All such operators have continuum counterparts, $\Box_\rho$, obtained by
averaging the discrete operators over all sprinklings of a given spacetime.
As anticipated in the introduction, the non-locality of the continuum
operators is parametrised by a length scale $l_k=\rho^{-1/d}$, taken
to be much larger than the discreteness scale $l$ in order to damp fluctuations. This
new mesoscopic scale could provide 
interesting phenomenology, since the non-locality of the d'Alembertian would survive at 
scales in which the continuum description of spacetime is approximately valid, hence the
continuum non-local d'Alembertian would describe the dynamics of a scalar field more accurately. 
Only at much smaller scales, of order $l$, would the discrete description then be necessary. 
Throughout the rest of this section we follow the notation of \cite{Glaser:2013sf}.

The minimal retarded non-local d'Alembertian in $d$-dimensional Minkowski spacetime $\mathbb{M}^d$
is given by~\footnote{Minimal here refers to the fact that  $N_d$ is the smallest integer in $d$-dimensions 
such that the continuum limit can be recovered, i.e. $\lim_{\rho\rightarrow\infty}\Box_{\rho}\phi(x)=\Box\phi(x)$,
e.g. $N_2=2$ and $N_4=3$. For details see \cite{Aslanbeigi:2014tg}.} 
\be
\Box_{\rho}^{(d)}\phi(x) = \rho^{\fr{2}{d}}\left(\alpha^{(d)}\phi(x) +\rho\, \beta^{(d)}\sum_{n=0}^{N_d}C_n^{(d)}\int_{J^-(x)}d^dy \,\fr{(\rho V(x,y))^n}{n!}e^{-\rho V(x,y)}\phi(y)\right),
\label{boxd}
\ee
where $N_{d}$ is a dimension dependent positive integer, $\rho=1/l_k^d$ and the coefficients $\alpha^{(d)}$, $\beta^{(d)}$ and $C_n^{(d)}$ can be found in Equations (12)-(15) of \cite{Glaser:2013sf}. 
This can be rewritten as
\be
\Box_{\rho}^{(d)}\phi(x) = \int_{J^-(x)}d^dy \,K^{(d)}(x,y)\phi(y),
\ee
where
\be
K^{(d)}(x,y):= \rho^{\fr{2}{d}}\left(\alpha^{(d)}\delta(x,y) +\rho\, \beta^{(d)}\sum_{n=0}^{N_d}C_n^{(d)}\fr{(\rho V(x,y))^n}{n!}e^{-\rho V(x,y)}\right).
\ee
Translational symmetry of $\mathbb{M}^d$ implies that $K^{(d)}(x,y)=K^{(d)}(x-y)$; therefore, 
letting $w=y-x$,
\begin{align}
\Box_{\rho}^{(d)}\phi(x) &= \int_{J^-(0)}d^dw \,K^{(d)}(-w)e^{w\cdot\del_x}\phi(x)=f^{(d)}(-\Box)\phi(x),
\end{align}
where we used the fact that since $K^{(d)}$ is Lorentz invariant (LI), it is function of $w^2:=w\cdot w$, and we defined
\be
f^{(d)}(-\Box):=\int_{J^-(0)} d^dw \,K^{(d)}(-w)e^{w\cdot\del_x}.
\label{fbox}
\ee 
The non-local equations of motion of a massless field, $\phi(x)$, living in $d$-dimensional Minkowski spacetime 
can therefore be written as
\be
f^{(d)}(-\Box)\phi(x) = 0.
\label{nleom}
\ee
The action of $f(-\Box)$ on $\phi$ can be defined in different ways, depending on the analytic 
properties of $f$: If $f(z)$ is everywhere analytic then we can represent it as the convergent 
power series expansion
\be
f(z)=\sum_{n=0}^{\infty}a_n z^n.
\label{fmom}
\ee 
Otherwise we may define it through its action on Laplace transforms, which
is how we will proceed.

The Laplace transform of $f(-\Box)$ in $d$-dimensions is given by \cite{Aslanbeigi:2014tg}
\be
f^{(d)}(k^2)=\rho^{2/d}\left(\alpha^{(d)}+\beta^{(d)}2(2\pi)^{d/2-1}Z^{\fr{2-d}{4}}\sum_{n=0}^{N_d}\fr{C_n^{(d)}}{n!}\gamma_d^n
\int_0^{\infty}ds\,s^{d(n+1/2)}e^{-\gamma_d s^d}K_{\fr{d}{2}-1}(Z^{1/2}s)\right),
\label{dalembFT}
\ee
where 
\be
Z=\fr{k\cdot k}{\rho^{2/d}},\quad\qquad \gamma_d=\fr{(\fr{\pi}{4})^{\fr{d-1}{2}}}{d\,\Gamma\left(\fr{d+1}{2}\right)},
\ee
and $K_{\nu}$ is the modified Bessel function of the second kind containing a cut along the negative real axis and 
for which we assume the principal value. Plane wave solutions $e^{ik\cdot x}$ to $f(-\Box)\phi=0$ lie in the kernel of $f(k^2)$. 
Aslanbeigi {\it et al.} \cite{Aslanbeigi:2014tg} showed that in 2 and 4 dimensions $f^{(d)}(k^2)=0$ iff $k^2=0$ and
$k^2=0, \zeta_4, \zeta^*_4$ respectively,
where $\Re(\zeta_4)<0$ and $\Im(\zeta_4)>0$ and the existence of the complex mass poles was 
determined numerically. Applying their analysis to other dimensions we obtain Table \ref{table:massless} 
for $2\le d\le7$
\begin{table}[h]
\caption{Roots of $f(z)$}
\centering
\scalebox{0.925}{
\begin{tabular}{ |c|c|c|c|c|c|c| } 
\hline
 $\;d=2\;$ & $\;d=3\;$ & $d=4$ & $d=5$ & $d=6$ & $d=7$ \\
\hline
$0$ & $0$ & $\;0,\zeta_4,\zeta^*_4\;$ & $\;0,\zeta_5,\zeta^*_5\;$ & $\;0,\zeta_6,\zeta^*_6,\eta_6,\eta^*_6\;$
& $\;0,\zeta_7,\zeta^*_7,\eta_7,\eta^*_7\;$\\
\hline
\end{tabular}
}
\label{table:massless}
\end{table}

\noindent where  $\Re(\eta_d)<0$ and $\Im(\eta_d)>0$, $d=4,\dots,7$.
\footnote{It should be noted that for $d\ge4$ the roots of $f^{(d)}$ 
away from the origin were found numerically.} 
A simple dimensional argument, confirmed by numerics, suggests that
$\zeta_d,\eta_d\propto \rho^{2/d}$ as is 
expected given that $\rho$ is the only dimensionful
parameter in the theory. The table above suggests that the number of zeros of $f^{(d)}$
grows with the number of dimensions, 
with a new pair appearing every time the dimensionality of the
spacetime goes up by two. Assuming this pattern continues for all $d$, then  
the number of zeros of the minimal d'Alembertian in $d$-dimensions would be 
$(d-1)$ for $d$ even, and $(d-2)$ for $d$ odd. 
Note that the complex mass solutions appear in complex conjugate pairs ensuring that the 
theory is CPT invariant -- the pole structure 
of the propagator in the complex $k^0$-plane has to be symmetric about the real
axis. This is expected since our action (see Equation~\ref{freelagrangian} below) 
is CPT-invariant: C being trivial for a real scalar field theory, and PT because
the non-local d'Alembertian $\Box_\rho$ is a function of the spacetime volume only, which
is PT-invariant in Minkowski spacetime.

We will see in Section \ref{FSNLQFT} that in order to establish the number of initial conditions required to specify a state of the system
(equivalently the number of propagating degrees of freedom), one
needs to know the degree of the roots of $f(z)$. Consider therefore the case $d=4$, which we will be concerned with
in Section \ref{4dim}. In this case there exist three roots: $z=0,\zeta_4,\zeta_4^*$. 
The root at the origin, $z=0$, is a branch point whose degree can be easily established to be 
$<1$, since $\lim_{z\rightarrow0}zf^{-1}(z) = 0$. Had we known the exact location of roots $z=\zeta_4,\zeta_4^*$, 
then a similar analysis would have given their degree. 
However, since we are only able to establish their existence numerically, we have 
had to resort to numerics to compute their degree. The numerical analysis is given in Appendix \ref{appC}, and 
suggests that the roots at $\pm\Omega,\pm{\Omega^*}$, 
where $\Omega:=k^0= \sqrt{\bold{k}^2-\zeta_4}$ and  $\Omega^*= \sqrt{\bold{k}^2-\zeta^*_4}$, are of order 1.
We will therefore assume this to be the case throughout the rest of this article.
\footnote{If the poles are of higher order, then the coefficients in the field expansion will
have an explicit time dependence.
}

It is interesting to note that the (formal) infrared expansion of (\ref{fbox})
in 2 and 4 dimensions
is given by~\footnote{The derivation of these expansions, together with the full series, is given in Appendix \ref{appA}}
\be
f^{(2)}(-\Box)= \Box-\frac{\Box^{2}}{2\rho}\left[\gamma+\ln\left(\frac{-\Box}{2\rho}\right)\right]+\dots,
\label{2dEx}
\ee
and
\be
f^{(4)}(-\Box)= \Box-\frac{3}{2\pi\sqrt{6}}\frac{\Box^{2}}{\sqrt{\rho}}\left[3\gamma-2+\ln\left(\frac{3\Box^{2}}{2\pi\rho}\right)\right]+\dots,
\label{4dEx}
\ee
respectively (here $\gamma$ is Euler-Mascheroni's constant).~\footnote{We would like to point out that
during the final stages of writing up, an article by S. Johnston \cite{Johnston:2014sf} appeared in which he obtains the same 
corrections as above, and the same power series 
expressions found in Appendix \ref{appA}.} 
Therefore the non-locality
is manifest even in the first order corrections to the standard continuum d'Alembertian in the IR limit.

\section{Massive Extension: Nonlocal Klein-Gordon Equation
\label{massive}}

In this section we generalise the nonlocal wave equation (\ref{nleom}) to include massive fields.

Consider the following naive massive extension to (\ref{nleom})
\be
\left[f(-\Box) - m^2\right]\phi(x)=0,
\label{naive}
\ee
where $m^2\in \mathbb{R}^+$. This equation does not admit plane wave solutions
(for finite $\rho$) with $k^2=-m^2$, since
$f(k^2)\in\mathbb{C}$ if $k^2<0$ (although it does reduce to the standard Klein-Gordon equation in the limit
$\rho\rightarrow\infty$).
Instead, we define the equation of motion of the massive field to be
\be
f(-\Box+m^2)\phi(x)=0.
\label{fmass}
\ee
We call this the non-local KG equation.
Unlike Equation (\ref{naive}), this does admit plane wave solutions with massive dispersion relations $k^2=-m^2$, 
as well as possessing the correct continuum limit: 
\be
\lim_{\rho\rightarrow\infty} f(-\Box+m^2)\phi(x) = (\Box+m^2)\phi(x).
\ee
Indeed, from Section \ref{NLBOX} we know that 
(ignoring, if any, the complex mass solutions which will simply undergo a 
translation along the real axis) $f(z)=0$ if $z=0$. Hence, in momentum space, we have that
$f(k^2+m^2)=0$ if $k^2=-m^2$ for any $\rho$. Note that the functional form of $f$ has not changed
so the branch cut
remains but with the branch point now shifted to $k^2=-m^2$.
\footnote{An interesting problem for the causal set community is to determine the inverse Laplace transform of this function,
which would correspond to the massive version of Equation (\ref{boxd}) in position space.
This would lead to a definition of the KG equation
on a causal set.\label{footmass}}
Table \ref{table:massless} can therefore be trivially extended to include the massive case to give
\begin{table}[h]
\caption{Roots of $f(z+m^2)$}
\centering
\scalebox{0.925}{
\begin{tabular}{ |c|c|c|c|c|c|c| } 
\hline
 $\;d=2\;$ & $\;d=3\;$ & $d=4$ & $d=5$ & $d=6$ & $d=7$ \\
\hline
$-m^2$ & $-m^2$ & $\;-m^2,\tilde{\zeta}_4,\tilde{\zeta}^*_4\;$ & $\;-m^2,\tilde{\zeta}_5,\tilde{\zeta}^*_5\;$ & $\;-m^2,
\tilde{\zeta}_6,\tilde{\zeta}^*_6,\tilde{\eta}_6,\tilde{\eta}^*_6\;$
& $\;-m^2,\tilde{\zeta}_7,\tilde{\zeta}^*_7,\tilde{\eta}_7,\tilde{\eta}^*_7\;$\\
\hline
\end{tabular}
}
\label{table:massive}
\end{table}

\noindent where $\tilde{\zeta}_d=\zeta_d -m^2$ and $\tilde{\eta}_d=\eta_d -m^2$.

\section{Huygens' Principle and the Nonlocal d'Alembertians
\label{huygens}}

Huygens' principle (HP) can be interpreted as stating that in spacetime dimensions $d=2n+2$, $n>0$, the propagators
of the wave equation $\Box G =\delta$ have support on the light-cone, while in dimensions $d=2n+1$ they also have support
{\it inside} the light-cone. The case $d=2$ is degenerate and somewhat counter intuitive since the Green functions are constant
inside the light cone.  

\begin{figure}[h]
\begin{center}
\includegraphics[scale=0.35]{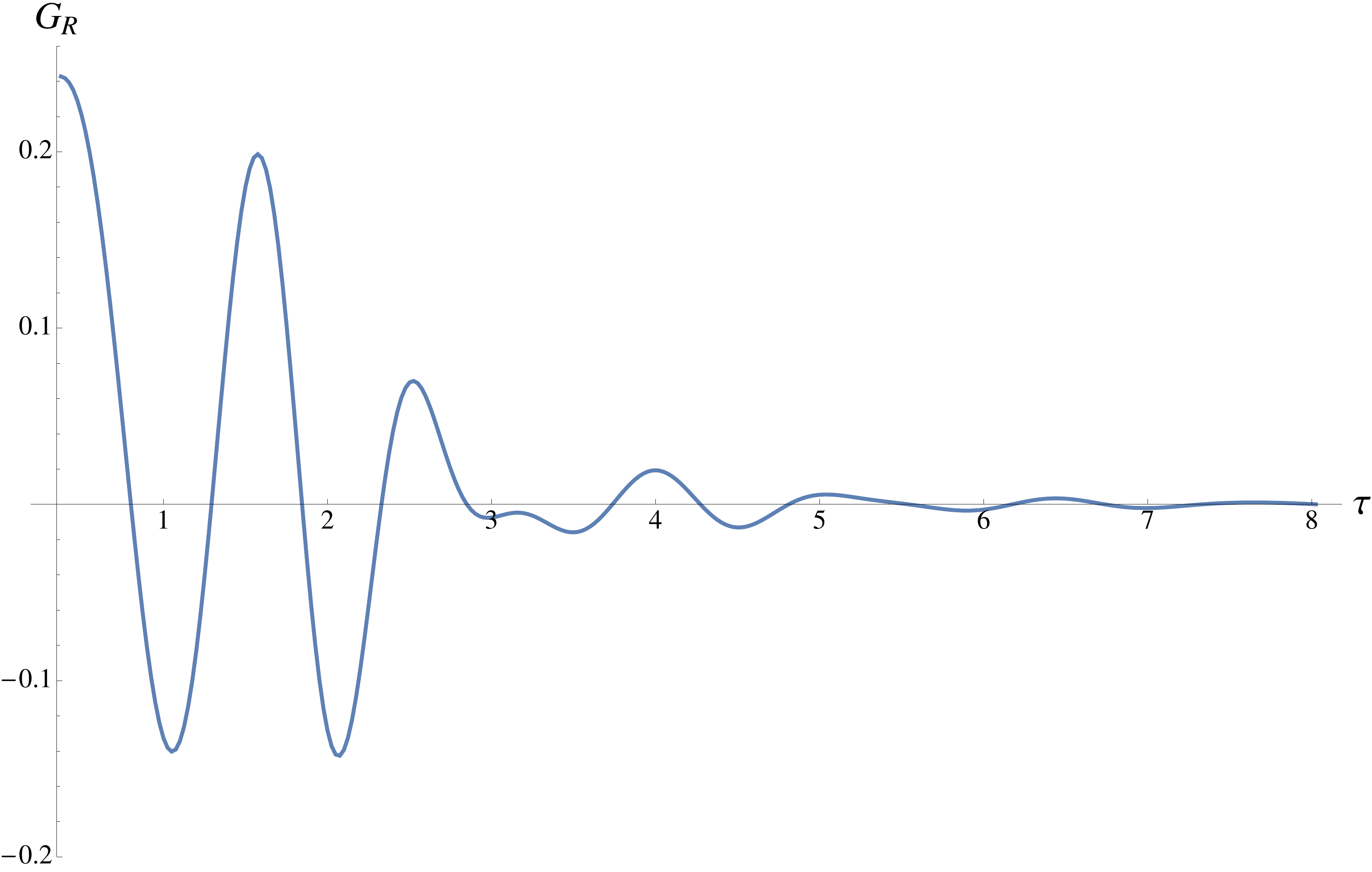}
\caption{4 dimensional retarded, non-local Green function as a function of proper time $\tau$ for $\rho=1$. Note
how the function has non-zero support inside the light cone ($\tau>0$) but decreases rapidly and asymptotes 0 
in the limit $\tau/\rho\rightarrow\infty$. For small $\tau\sim O(\rho)$ we therefore have large deviations from the
local retarded Green function in 4$d$, but recover the usual result in the limit.}
\label{4dleak}
\end{center}
\end{figure}

It is interesting to ask therefore if such a principle also applies in the case where $\Box$ is replaced by $f(-\Box)$. A priori
there is no reason why it should, since the analytic properties of $f^{-1}(-\Box)$ are in general very different from those
of $\Box^{-1}$. Indeed, using the explicit form for the retarded Green function in $d$-dimensions given in \cite{Aslanbeigi:2014tg},
it is possible to show that for $d=4$ the retarded Green function has support inside the light cone. Explicitly the
$d$-dimensional Green function is given by
\begin{align}
   \rho^{2/d}G(x-y)
   &\overset{x\succ y}{=}
    \frac{2(-1)^{1+\frac{d}{2}}\tau_{xy}^{1-\frac{d}{2}}}{(2\pi)^{d/2}}\rho^{2/d}
    \int_0^{\infty}d\xi~\xi^{d/2}
     \frac{\text{Im}\left[ f(-\xi^2+i\epsilon)\right]}{\left| f(-\xi^2+i\epsilon)\right|^2}J_{\frac{d}{2}-1}(\tau_{xy}\xi),
     \label{green}
\end{align}
where $J_n(z)$ is the Bessel function of the first kind, and $x\succ y$ means $x$ lies to the future of $y$.
A numerical plot of this function against proper time $\tau$ is given in Figure~\ref{4dleak}.
\footnote{In the plot for $d=2$ we have corrected the expression for the Green function by an overall factor
of 2. This is because Equation (\ref{green}) for $d=2$ was tending to -1/4 in the limit
$\tau/\rho\rightarrow\infty$,  rather than -1/2, which is the value that the retarded Green function of $\Box$
takes inside the light cone in 2$d$. We therefore believe that (at least in the 2 dimensional case), Equation (\ref{green})
is off by an overall factor of 2.} 
Note how
the function is non vanishing for nonzero values of proper time but decays to zero rapidly for $\tau\gg \rho^{-1/d}$
(in the plot $\rho=1$).
This is to be expected since the Green function (\ref{green}) reduces to the standard, local Green function
which has no support inside the light cone, in the limit $\rho\rightarrow\infty$. 
This plot therefore demonstrates the failure of HP in 4$d$. Since 
all other even dimensional Green functions are similar in form one should expect this to be true for all
$d=2n+4$, $n\ge0$.
\begin{figure}[h]
\begin{center}
\includegraphics[scale=0.385]{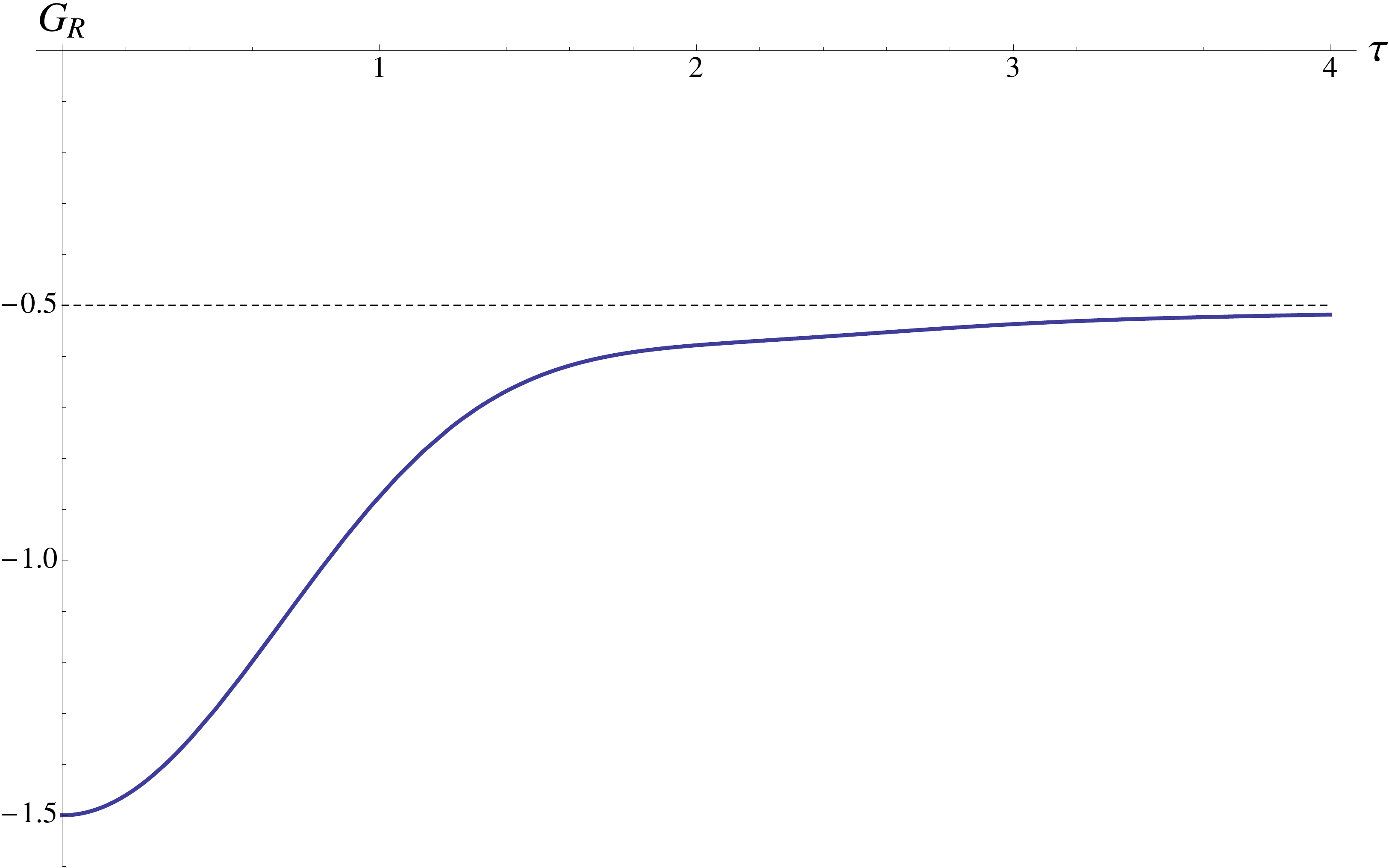}
\caption{2 dimensional retarded, non-local Green function as a function of proper time $\tau$ for $\rho=1$. Note how for
$\tau\sim O(\rho)$ the non-local retarded Green function deviates from the usual local one, but asymptotes the local Green function
in the limit $\tau/\rho\rightarrow\infty$.}
\label{2dleak}
\end{center}
\end{figure}
Even though HP does not strictly apply in $d=2$, it is nonetheless interesting to see how the nonlocal Green function
is modified in this case. The plot of $G_R$ against proper time $\tau$ is shown in Figure~\ref{2dleak}. 
As one can see
the function is no longer constant for all $\tau>0$ but, crucially, it tends to a constant in the limit $\tau/\rho\rightarrow\infty$.

The above observations, though purely classical at this point, are important for the quantum theory. Indeed, the Pauli-Jordan 
function -- defined as the vacuum expectation value of the commutator of fields -- is given by $i$ times the difference between
the retarded and advanced propagators. Hence for a 4$d$ theory based on such non-local dynamics, it will fail to vanish
inside the light-cone, unlike a QFT based on local dynamics given by just $\Box$. Although we will not investigate it in this article,
it would be interesting to apply the Johnston-Sorkin construction of QFT \cite{Afshordi:2012ij}, which is based on the Pauli-Jordan function
and therefore on the retarded and advanced propagators, to these non-local theories and compare this with the results of this 
article. This might shed light on the Sweety-Salty duality referred to in Section 3.2 of \cite{Johnston:2014sf}.

\section{Free Scalar Nonlocal QFT}
\label{FSNLQFT}

In the following sections we construct free scalar quantum field theories based on the nonlocal dynamics defined
by $f(-\Box)$. We closely follow the formalism set out in \cite{Barci:1996kq} for canonically quantising nonlocal field
equations. Everything we say can be extended to the massive case by simply replacing $f(-\Box)$ by $f(-\Box+m^2)$.
\footnote{Although we can extend the definition of the non-local dynamics to include massive fields in the continuum,
it is not clear what the corresponding theory on the causal set is. As such this extension represents a phenomenological
model rather than a model derived directly from the causal set. It remains an interesting open issue as to how one should
include a massive term in the causal set in order to produce continuum equations of motion of the type $f(-\Box+m^2)\phi=0$
(c.f. Footnote~\ref{footmass} in Section~\ref{massive})}
We now outline this formalism.

We begin with the (free) non-local Lagrangian
\footnote{Note that 
this kind of non-locality does not imply a violation of micro-causality. 
As such it is fundamentally different from the kind of non-locality often considered in QFT, e.g. \cite{Marolf:2014yu}.
}
\be
\mathcal{L}= \phi(x)f(-\Box)\phi(x),
\label{freelagrangian}
\ee
which leads to the following Euler-Lagrange equations (see Section 2 of \cite{Barci:1996kq} 
and references therein)
\footnote{It is interesting to note that the variational principle applied to non-local actions 
always leads to acausal equations of motion \cite{Jaccard:2013fk}.}
\be
f(-\Box)\phi(x)=0.
\label{eom}
\ee 
A general solution to the equations of motion can be written as
\cite{Barci:qf}
\be
\phi(x) = \int_{\Gamma}dk^0\int d^{d-1}\kvec\,\fr{a(k)}{f(k^2)}e^{ik\cdot x},
\label{solution}
\ee
where $\Gamma$ runs from $-\infty$ to $\infty$ for $\Im(k^0)> s$ and $\infty$ to $-\infty$ for $\Im(k)<-s$ for
$f^{-1}(k^2)$ analytic in $\{k^0\in \mathbb{C}\;|\; |\Im(k^0)|> s\}$, and $a(k)$ is an entire analytic function
(note that for $a(k)=1$ this contour defines the usual Pauli-Jordan function, which is in the kernel 
of the d'Alembertian).
\footnote{$f^{-1}(z)/z^s$ also has to be bounded continuous in $\{k\in \mathbb{C}\;|\; |\Im(k)|\ge s\}$, see \cite{Barci:qf}.}
By continuously deforming the path $\Gamma$ around the singularities of $f^{-1}$, Equation (\ref{solution}) 
can be rewritten as
\be
\phi(x) = \int d^dk\, \theta(k^0)\Delta(f^{-1})(a(k)e^{ik\cdot x}-a^*(k)e^{-ik\cdot x}) 
+ \sum_{i=1}^N\int_{\Gamma_i}d^dk\,\fr{a(k)}{f(k^2)}e^{ik\cdot x},
\label{solution2}
\ee
where we used reality conditions on $\phi$ to determine that $a(-k)={a}(k)^*$,
and we have defined the discontinuity functional across the branch cut in $f^{-1}$ to be
\be
\Delta(f^{-1}) := f^{-1}(-(k^0+i\epsilon)^2+k^2)-f^{-1}(-(k^0-i\epsilon)^2+k^2),
\ee
and the $\Gamma_i$'s are loops surrounding isolated singularities of $f^{-1}$. It is interesting to note that the general
solution (\ref{solution2}), when $f^{-1}(z)$ contains a branch cut (as is the case here), is {\it not} a linear superposition of 
plane waves with dispersion relation $k^2=0$ only. Rather, the functional $i\Delta$ in general gives non-zero weight 
to all $k^2=-m^2$ with $m^2\ge0$, hence $\phi(x)$ is a superposition of all possible (massless and massive) plane waves.
Nonetheless (ignoring the existence of complex mass solutions for the time being), 
it would not be correct to state that plane waves with $k^2<0$ are part of the basis of the solution space, since
they themselves do not satisfy the equations of motion. It is also incorrect to view plane waves with $k^2=0$ as a 
basis since the expansion of a general solution clearly also contains plane waves with $k^2\ne0$.

Naively one might think that the infinite number of derivatives present in the equations of motion requires
having to specify an infinite number of initial conditions to determine a solution. However, it can be shown that a large class
of infinite order differential equations admit a well-defined initial value problem \cite{Barnaby:2008ud}, requiring only a finite
number of initial conditions; the precise number depending on the number of poles in the propagator and their 
degree. 
Equivalently, poles in the propagator can be used to determine the number of degrees of freedom; with every pole
of degree $\le1$ representing a single degree of freedom. 
So, for example, in 4-dimensions where the number of roots of $f(k^2)$ is 3, one needs 6 initial conditions
to specify a solution: two for the pole at $k^2=0$ and two for each of the complex mass roots $k^2=\zeta_4,\zeta^*_4$.
As such one should expect 3 propagating degrees of freedom. 
In the quantum theory these degrees of freedom are
quantised and appear as states in the physical Hilbert space of the theory. Thus, coming back to our 4$d$ example,
we expect the Hilbert space of the (free) quantum theory to contain states associated to the 3 poles in the propagator. 
In fact, as we will see in Section \ref{4dim}, it is possible to construct the quantum theory such that 
states associated to the complex mass 
modes do not appear asymptotically -- in a sense, they behave as if propagating in a medium that acts as a perfect absorber --
while the massless ones do, as is suggested by the above argument. In Appendix \ref{appB} we explicitly check that the continuum
of massive modes associated to the cut $k^2<0$ do not appear in the asymptotic states of the quantum theory.\\
\indent 
In \cite{do1992canonical}, the canonical
quantisation of theories containing fractional powers of the d'Alembertian was performed. Although ultimately successful,
this method is not without difficulties, having an infinite set of second class constraints to solve and ill-defined 
Poisson brackets between the conjugate variables. 
We therefore quantise our system following D. Barci {\it et. al}  \cite{Barci:1996kq}. 
Their method consists in observing that Schwinger's quantisation method
implies
the Hamiltonian must be the generator of time evolution at the level of quantum theory. 
This is equivalent to imposing Heisenberg's equations of motion on the quantum field: $\dot{\phi} = i[H,\phi]$.
\footnote{Although this quantisation procedure is not canonical, Barci and Oxman found that when applied to theories
containing fractional powers of the d'Alembertian, their results were consistent with those of \cite{do1992canonical}.}
The definition of a vacuum state proceeds as usual for the massless
modes associated to the branch point at $k^2=0$, since
their creation and annihilation operators possess canonical commutation relations. As for 
the complex mass modes, their creation and annihilation operators possess peculiar commutation relations (akin
to those between $q$ and $p$ in standard quantum mechanics) and therefore require further care in defining the vacuum
state and the corresponding Fock space. This will be analysed in Subsection \ref{4dim} below.

In general dimension, $d$, the Hamiltonian can be shown to be given by (c.f. \cite{bollini1987lagrangian})
\be
H=-\int d^{d-1}\bold{x}\int d^dk\int d^dk'\, e^{i(k+k')\cdot x}\phi(k)\phi(k')k'^0\fr{k'^0-k^0}{k^2-k'^2}(f(k^2)-f(k'^2)) + \int d^{d-1}\bold{x}\,\mathcal{L},
\label{Ham}
\ee
where the last term vanishes when the field is on-shell.

\subsection{2 Dimensions}

In 2 dimensions the momentum space d'Alembertian can be expressed in the relatively simple form
\be
\label{2dFT}
f(k^2)=-k^2e^{k^2/2\rho}\,E_2(k^2/2\rho),
\ee
where $E_2$ is the Exponential Integral of the second kind:
\be
E_2(z) := ze^{-z}\int_0^{\infty}\fr{e^{-t}}{(t+z)^2}dt,
\ee
containing a branch cut along $z\le0$ \cite{gradshteyn2009tables}. 
\begin{figure}
\begin{center}
\includegraphics[scale=0.35]{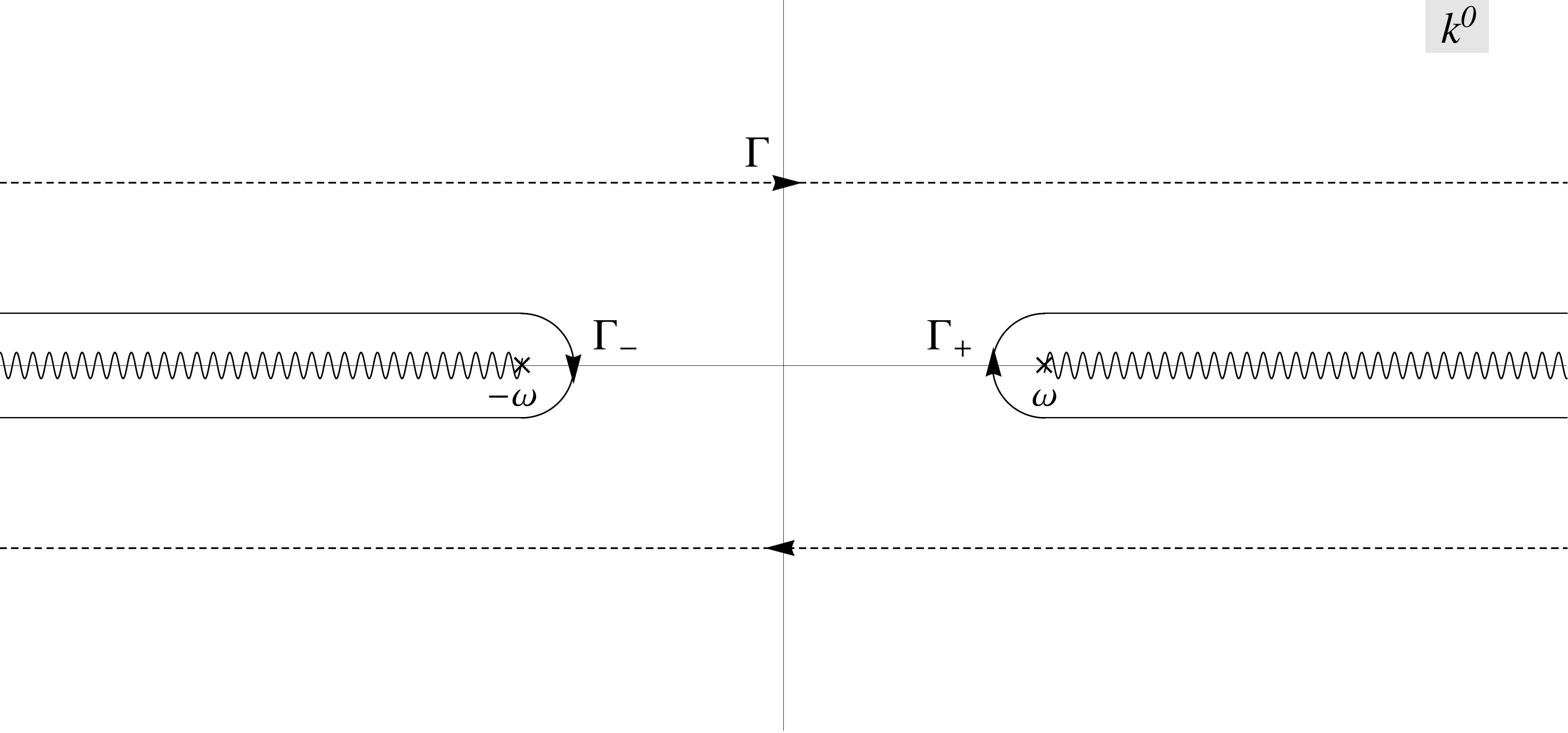}
\caption{Pole structure of $f^{-1}(k^2)$ in 2$d$ in the complex $k^0$ plane, together with a choice of contour, $\Gamma$, which gives a solution
to the equations of motion. The continuous paths represent a deformation of $\Gamma$: $\Gamma=\Gamma_+\cup\Gamma_-$.}
\label{contour2d}
\end{center}
\end{figure}

\begin{figure}
\begin{center}
\includegraphics[scale=0.35]{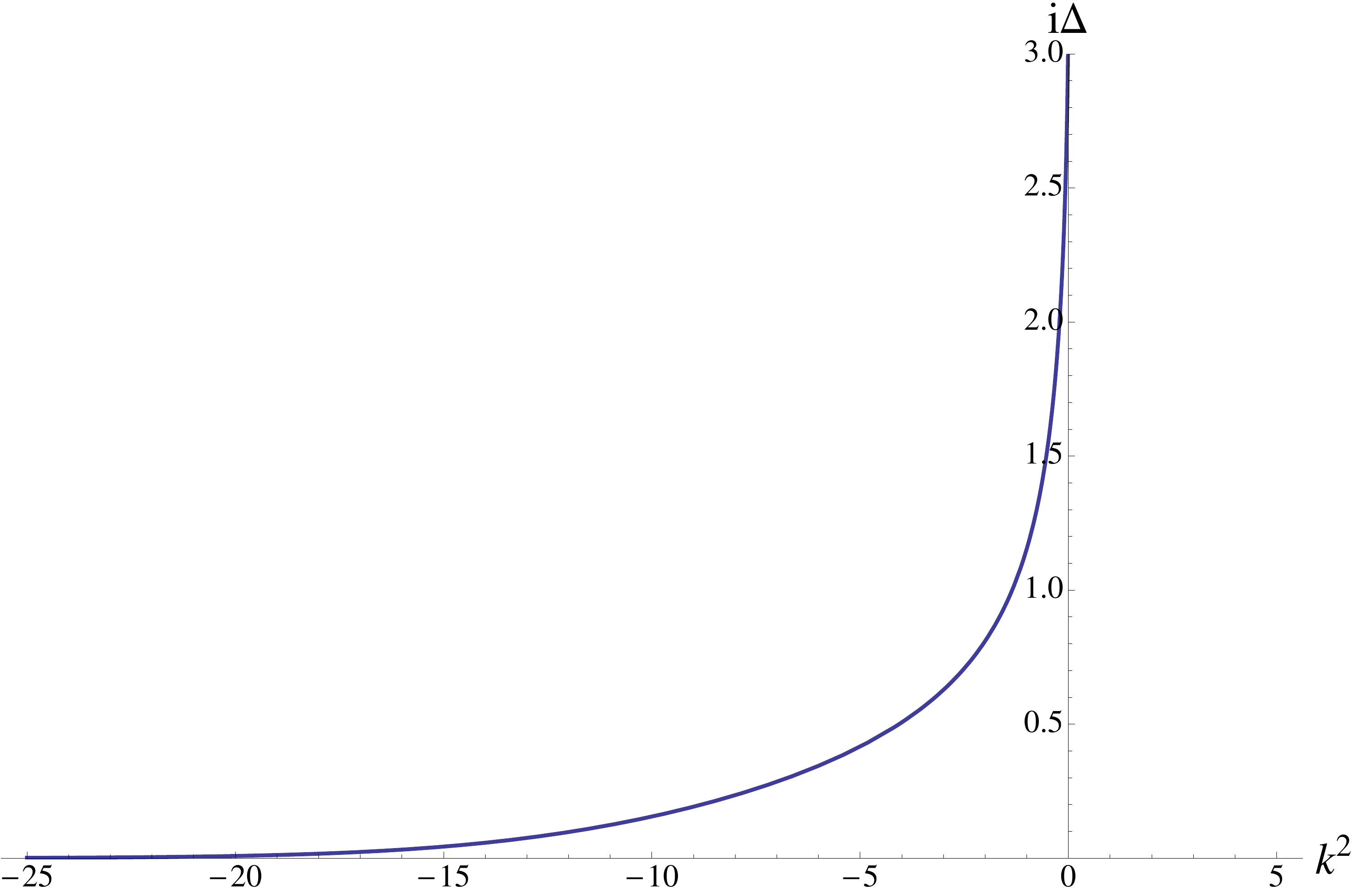}
\caption{Plot of $i\Delta(f^{-1})$ as a function of $k^2$ for $k^0>0$ and $\rho=1$. Note how the function blows up at the origin, ensuring that
massless modes provide the greatest contribution, and is non zero for all $k^2<0$ but rapidly decays to zero.}
\label{fig:discfunc2d}
\end{center}
\end{figure}

To find the general solution (Equation (\ref{solution})), we note that as a function of $k^0\in \mathbb{C}$,
$f^{-1}$ is analytic for $\Im(k^0)>0$, hence we can deform the contour $\Gamma$
(see Figure~\ref{contour2d}) such that 
\be
\phi(x) = \int_{-\infty}^{\infty}d^2k\,\theta(k^0)\Delta(f^{-1})\left(a(k)e^{ik\cdot x}-a(k)^*e^{-ik\cdot x}\right).
\label{soln2d}
\ee
After some manipulations the discontinuity functional for $k^0>0$ can be rewritten as
\begin{align}
i\Delta(f^{-1})
&=\lim_{\epsilon\rightarrow0}\fr{2\,e^{-k^2/2\rho}}{k^2} \fr{\Im[E_2((k^2+i\epsilon)/2\rho)]}{|E_2((k^2+i\epsilon)/2\rho)|^2}.
\label{discfunc2d}
\end{align}
Figure~\ref{fig:discfunc2d} shows a plot of (\ref{discfunc2d}) for $\rho= 1$.
Note that all modes have positive weight, and hence positive energy with respect to the Hamiltonian defined below (see Equation
\ref{nham}).
Consistency with $\Box_{\rho}\rightarrow\Box$ as $\rho\rightarrow\infty$
implies that 
$i\Delta(k^{2}) \rightarrow \delta(k^2)$ in the limit $\rho\rightarrow\infty$. Indeed one can show 
\begin{align}
\lim_{\rho\rightarrow\infty}\Delta(z)=&
\lim_{\epsilon\rightarrow 0}\left\{-\frac{1}{z-i \epsilon }+\frac{1}{z+i \epsilon }+\frac{i}{\rho }\text{Im}
\left[\fr{z+i\epsilon}{z-i\epsilon}\left(\ln\left(\fr{z+i\epsilon}{2\rho}\right)+\gamma\right)
\right]
+O(\rho^{-2})\right\}.
\end{align}
The zeroth order term is just a representation of the delta function (i.e. $-2\pi i\delta(z)$),
hence we recover the standard local theory in the limit $\rho\rightarrow\infty$. 
The second term represents the first order correction in the $1/\rho$ series expansion, and
it is interesting to note that this term already contains a branch cut, and therefore 
implies the presence of a continuum of massive modes.

Substituting (\ref{soln2d}) into (\ref{Ham}) and integrating over $x$ and $k'^1$  we find the on-shell Hamiltonian
\be
H=-\int dk\int_{\Gamma} dk^0\int_{\Gamma'} dk'^0\, e^{-i(k^0+k'^0)t}\fr{k'^0}{k'^0+k^0}
a(k^0,k)a(k'^0,-k)(f^{-1}(k'^0,-k)-f^{-1}(k^0,k)).
\label{intham}
\ee
Without loss of generality, the contour $\Gamma'$ is chosen such that it runs above $\Gamma$ for $\Im(k^0)>0$ and
below for $\Im(k^0)<0$. It is then straightforward to show using Cauchy's residue theorem that
\be
H=2\pi \int_{-\infty}^{\infty} d^2k\,k^0\delta^+G(k)(a(k)a(k)^*+a(k)^*a(k)),
\ee
where $\delta^+G(k):=i\theta(k^0)\Delta(f^{-1})$.
Note that, unlike the local massless theory where only massless modes ($k^2=0$) appear in the Hamiltonian (i.e. where $\dg\sim\delta(k^2)$),
here all modes with $k^2\in \text{supp}(\dg)$ are present,
implying the existence of a continuum
of massive modes. This property is similar to that of interacting local QFTs 
(c.f. the Kallen-Lehman spectral representation), except that our theory contains these modes already at tree level and 
in the absence of any interactions. We will see later however 
that these modes are not present in the asymptotic states of the quantum theory 
\footnote{The term asymptotic comes from the definition of asymptotic field operators in 
interacting QFTs as defined by Greenberg \cite{Greenberg:1961if}}; which is consistent with the 
fact that the theory possesses a single propagating degree of freedom coming from the 
only singularity of the propagator at $k^2=0$.

The quantum theory is defined by promoting the $a$'s and $a^*$'s to operators:
$\hat{a}$ and $\hat{a}^{\dagger}$ respectively, and imposing Heisenberg's equations of motion. 
We find
\be
[H,a]=-k^0a,\qquad [H,a^{\dagger}]=k^0a^{\dagger}.
\label{ladder}
\ee
For $k^0>0$, $a(k)$ and $a(k)^{\dagger}$ correspond to raising and lowering operators respectively
(we have dropped hats to denote operators).
The vacuum is defined to be the state such that
\be
a(k)|0\rangle = 0 , \quad \forall k^2\in \text{supp}(\delta^+G), \;k^0>0.
\ee
Finally we define the normal ordered Hamiltonian
\be
:H: = 4\pi\int d^2k\; k^0\dg a(k^0,k)^{\dagger}a(k^0,k)
\label{nham}
\ee
for which $\vac$ is the zero energy eigenstate. Using this in (\ref{ladder}) we obtain the commutation relations
\begin{align}
[a(k),a(k')] = [a(k)^{\dagger},a(k')^{\dagger}] = 0, \qquad 4\pi \dg[a(k),a(k')^{\dagger}] =\delta^{(2)}(k-k')
\end{align}
for $k^2,\;k'^2\in \suppdg$.

Having defined the vacuum state we can now define the Wightman function,
\begin{align}
W(x-y)&:=\langle0| \phi(x)\phi(y) \vac\nn &= \fr{1}{4\pi}\int d^2k\, \dg e^{ik\cdot (x-y)}\\
&= \fr{i}{4\pi}\int_{\Gamma_+} d^2k\, f^{-1}(k^2)e^{ik\cdot (x-y)},
\end{align}
and the Feynman propagator of the theory,
\begin{align}
G_F(x-y)&:=\langle0| T\{\phi(x)\phi(y)\} \vac \nn
&= \theta(x^0-y^0)W(x-y)+\theta(y^0-x^0)W(y-x)\\
&= \fr{i}{4\pi}\int_{\Gamma_F} d^2k\, f^{-1}(k^2)e^{ik\cdot (x-y)},
\end{align}
where $\Gamma_F = \theta(x^0-y^0)\Gamma_+ - \theta(y^0-x^0)\Gamma_-$. 

\subsection{4 Dimensions\label{4dim}}

\begin{figure}
\begin{center}
\includegraphics[scale=0.35]{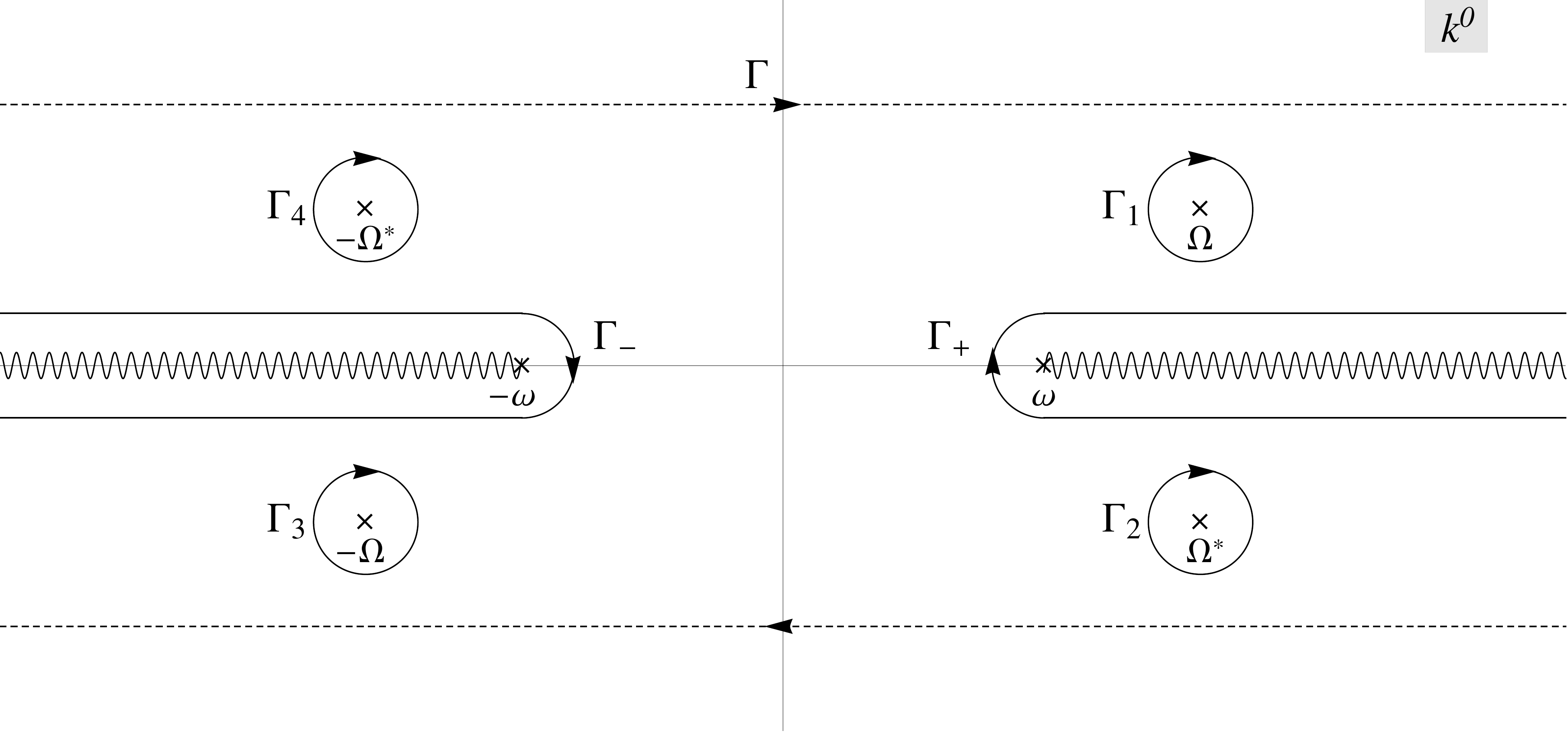}
\caption{Pole structure of $f^{-1}(k^2)$ in 4$d$ in the complex $k^0$ plane, together with a choice of contour, $\Gamma$, which gives a solution
to the equations of motion. The continuous paths represent a deformation of $\Gamma$, i.e. $\Gamma=\Gamma_+\cup\Gamma_-\cup_{i=1}^4\Gamma_i$.}
\label{contour4d}
\end{center}
\end{figure}

\begin{figure}
\begin{center}
\includegraphics[scale=0.35]{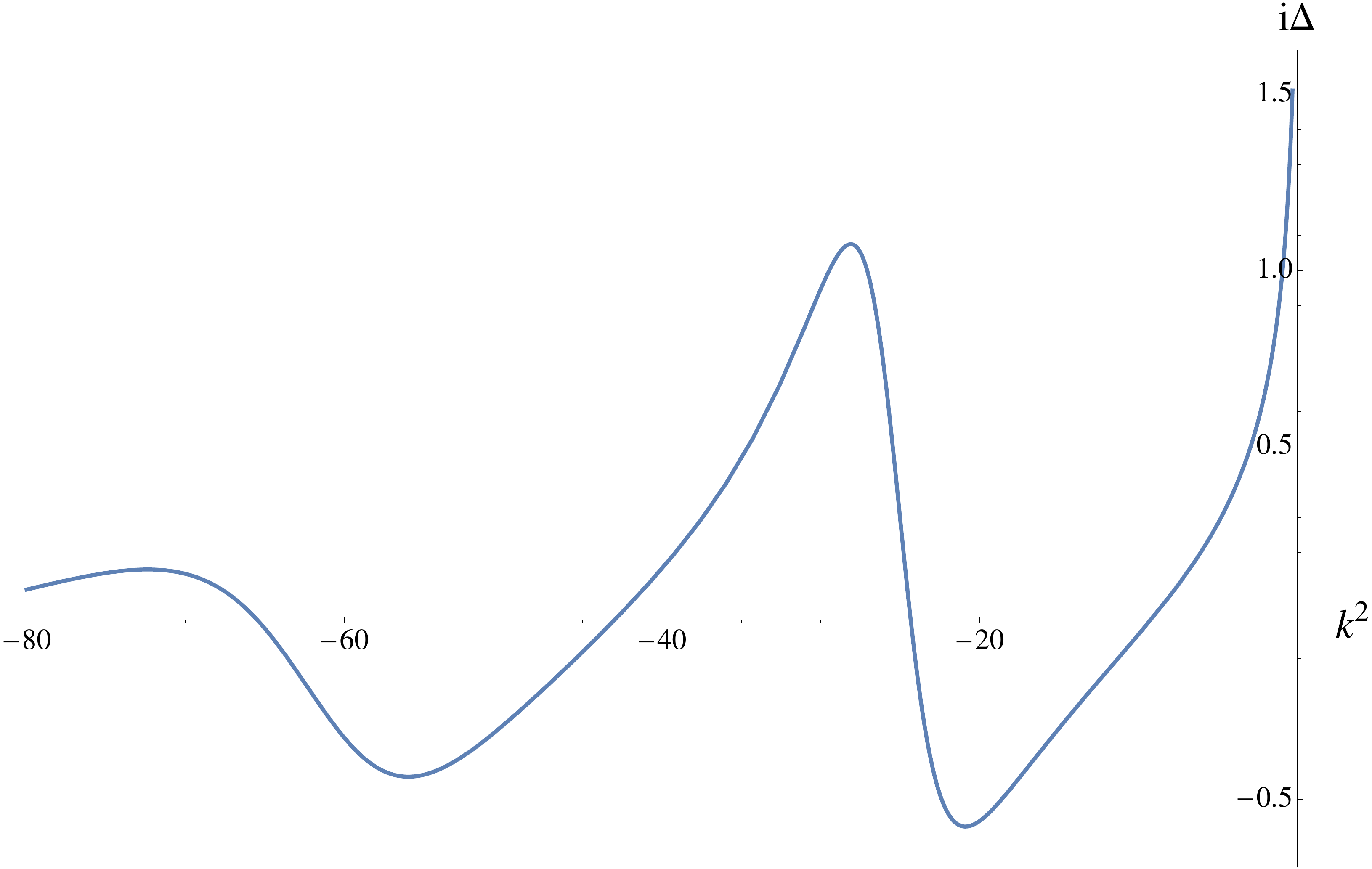}
\caption{Numerical plot of $i\Delta(f^{-1})$ as a function of $k^2$ for $\rho=1$. Note how the function blows up at the 
origin, ensuring that massless modes provide the dominant contribution, and rapidly decays in the limit 
$k^2/\sqrt{\rho}\rightarrow\infty$. Note also that, unlike the case $d=2$, this
functional fails to be positive definite for all $k^2<0$. 
}
\label{fig:DiscFunc4d}
\end{center}
\end{figure}

The 4 dimensional momentum space d'Alembertian is given by
\be
f^{(4)}(k^{2})=-\fr{4}{\sqrt{6}}\sqrt{\rho}+\fr{16}{\sqrt{6}}\pi\rho^{3/2}\int_0^{\infty} d\tau \,\fr{\tau^2}{\sqrt{k^{2}}}e^{-\rho V_4}
K_1(\tau\sqrt{k^{2}})\left(1-9\rho V_4+8\rho^2V_4^2-\fr{4}{3}\rho^3V_4^3\right),
\label{dalemb4d}
\ee
where $V_4=\pi\tau^4/24$ (from here on we will drop the superscript denoting the dimension we are working in). 
This function has zeros at $k^2=0,\zeta_4,\zeta^*_4$ and a branch cut along $k^2\le0$ (see Section \ref{NLBOX}), therefore
a general solution to (\ref{eom}) is 
\be
\phi(x) = \int d^4k\, \theta(k^0)\Delta(f^{-1})(a(k)e^{ik\cdot x}-a(k)^*e^{-ik\cdot x}) 
+ \sum_{i=1}^4\int_{\Gamma_i}d^4k\,\fr{a(k)}{f(k^2)}e^{ik\cdot x},
\label{soln4d}
\ee
where again $\Delta(f^{-1}) = f^{-1}(-(k^0+i\epsilon)^2+k^2)-f^{-1}(-(k^0-i\epsilon)^2+k^2)$ and the $\Gamma_i$
are loops surrounding the isolated singularities of $f^{-1}$ at $k^0=\pm\Omega,\pm\Omega^*$ (recall $\Omega= \sqrt{\bold{k}^2-\zeta_4}$),
obtained by continuously deforming the contour $\Gamma$, see Figure~\ref{contour4d}.
The discontinuity functional $i\Delta$
is shown in Figure~\ref{fig:DiscFunc4d}.
Note that unlike the 2$d$ case, this functional is {\it not} positive definite, implying that
some modes will have negative energy with respect to the non-local Hamiltonian defined below (Equation (\ref{ham4d})).

Assuming the complex poles of $f^{-1}$ are simple (see Section \ref{NLBOX} and Appendix \ref{appC}), 
we denote the residues of $f^{-1}$ by
\begin{align}
g_1(\Omega,\bold{k}) &= \text{Res}[f^{-1}(k^0,\bold{k}),\; k^0=\Omega],\\
g_2(\Omega^*,\bold{k}) &= \text{Res}[f^{-1}(k^0,\bold{k}),\; k^0=\Omega^*],
\end{align}
the other two residues being trivially related to these. 
Using Cauchy's residue theorem and imposing reality conditions we can then write (\ref{soln4d}) as
\begin{align}
\phi(x) = &\int d^4k\, \theta(k^0)\Delta(f^{-1})(a(k)e^{ik\cdot x}-a(k)^*e^{-ik\cdot x}) \nn
-2\pi i &\int d^3\bold{k}\,\left(g_1(\Omega,\bold{k})a(\Omega,\bold{k})e^{i\kappa\cdot x}
-\;g_1(\Omega,\bold{k})a(\Omega^*,\bold{k})^*e^{-i\kappa\cdot x}\right.\nn
&\qquad\;\left.+\;g_1(\Omega,\bold{k})^*a(\Omega^*,\bold{k})e^{i\kappa^*\cdot x}
-g_1(\Omega,\bold{k})^*a(\Omega,\bold{k})^*e^{-i\kappa^*\cdot x}\right),
\label{soln4d2}
\end{align}
where $\kappa=(\Omega,\bold{k})$ and  $\kappa^*=(\Omega^*,\bold{k})$.
Substituting this into (\ref{Ham}), integrating over $\bold{x}$ and $\kvec'$, and again choosing 
the contour $\Gamma'$ such that it runs above $\Gamma$ for $\Im(k^0)>0$ and
below for $\Im(k^0)<0$ we find the on-shell Hamiltonian
\begin{align}
H & =  2\pi \int d^4 k\,k^{0} \delta^{+}G(k)\left(a(k)a(k)^*+a(k)^*a(k)\right)\nn
&+4\pi^2\int d^3\bold{k}\, \Omega\, g_1(\Omega,\kvec) \{\aomega,a(\Omega^*,\bold{k})^*\} + \text{h.c.},
\label{ham4d}
\end{align}
where $\delta^+G(k)=i\theta(k^0)\Delta(f^{-1})$.

The quantum theory is defined by promoting the $a$'s and $a^*$'s to operators and by imposing Heisenberg's equations of motion: 
\begin{align}
[H,a(k)]&=-k^0a(k),\qquad [H,a(k)^{\dagger}]=k^0a(k)^{\dagger},\\
[H,c_{\kvec}]&=-\Omega \,c_{\kvec},\qquad\;\;\;\, [H,b_{\kvec}]=\Omega\, b_{\kvec},\\
[H,b^\dagger_{\kvec}]&=-\Omega^*\, b^\dagger_{\kvec},\qquad\;\;\;\,
[H,c^\dagger_{\kvec}]=\Omega^*\, c^\dagger_{\kvec},
\label{ladder4d}
\end{align}
where we defined $c_{\kvec}:=\aomega$,  $b_{\kvec}:=a(\Omega^*,\bold{k})^\dagger$,
and we have dropped hats to denote operators. 
For $k^0>0$, $a(k)$ and $a(k)^{\dagger}$ correspond to raising and lowering operators respectively; 
while the operators $c_{\kvec}$, $b_{\kvec}$ and their hermitian conjugates 
cannot be interpreted as raising and lowering operators for modes of energy
$\Omega$ and $\Omega^*$ respectively, as we will see shortly.

Substituting (\ref{ham4d}) into (\ref{ladder4d}) and defining $\beta^{-1}:=\Omega g_1(\Omega,\kvec)$ 
we find the set of commutation relations
\begin{align}
4\pi\dg[a(k),a(k')^{\dagger}]=\delta(k-k')&,\quad[a(k),a(k')]=[a(k)^{\dagger},a(k')^{\dagger}]=0,\\
[c_{\kvec},b_{\kvec'}]=\fr{\Omega\beta}{8\pi^2}\delta(\kvec-\kvec')&, \quad [c_{\kvec},c_{\kvec'}]=[b_{\kvec},b_{\kvec'}]= [c_{\kvec},b^\dagger_{\kvec'}]=\dots=0,
\label{4dCR}
\end{align}
with all commutators between the $a$'s and $b$'s or $c$'s vanishing.

From the commutation relations one can see that, unlike the 2$d$ theory, 
this theory has two distinct sectors: the bradyonic/luxonic sector corresponding to the creation and annihilation
operators $a(k)$ and $a(k)^{\dagger}$ respectively, and the complex mass sector corresponding to the $b_{\kvec}$, $c_{\kvec}$ and their Hermitean 
conjugates. In the former sector, which we refer to as the BL-sector (B for bradyons and L for luxons), the vacuum state $\vac$ is defined in the usual way
\be
a(k)|0\rangle = 0 , \quad \forall k^2\in \text{supp}(\delta^+G), \;k^0>0.
\ee
To define the vacuum state of the latter sector we follow the analysis of Bollini and Oxman \cite{Oxman:1992cm} closely.
A suitable representation is
\be
d_{\kvec}\rightarrow z,\quad d^\dagger_{\kvec}\rightarrow z^*,\quad
b_{\kvec}\rightarrow -i\fr{d}{dz},\quad b^\dagger_{\kvec}\rightarrow -i\fr{d}{dz^*},
\ee
for each $\kvec$, where $d_{\kvec}:=\fr{8\pi^2i}{\beta\Omega}c_{\kvec}$. The inner product defined to be
\be
\langle f|g\rangle= \int dz\int dz^* fg^*.
\ee
In this representation the complex-mass sector Hamiltonian becomes
\be
H=\int d^3\kvec -\Omega\left(z\fr{d}{dz}+\fr{1}{2}\right)+\Omega^*\left(z^*\fr{d}{dz^*}+\fr{1}{2}\right).
\ee
Similarly one can show that complex mass sector momentum-density operator is
\begin{equation}
p_{\textbf{k}}=\frac{i\textbf{k}}{2}\left\{z,-i\frac{\partial}{\partial z}\right\}+h.c.=\textbf{k}\left(z\frac{\partial}{\partial z}-z^*\frac{\partial}{\partial z^*}\right),
\end{equation}
so that its zero eigenvalue eigenfunctions are given by functions $f = f(zz^*)$. The vacuum state, $\vac$,  is defined to be the zero momentum eigenfunction of the Hamiltonian with zero energy, and can be shown to be given by
\be
f_0(zz^*) =\fr{1}{\sqrt{zz^*}}.
\ee
Note that in this sector the energy is {\it not} proportional to $\Omega$, nor $|\Omega|$. 
In fact, the eigenvalue equation for zero momentum eigenfunctions: $h_{\kvec}f_E(zz^*) = E f_E(zz^*)$, 
can be shown to have solutions (up to a normalisation factor)
%
%
\be
f_E=(zz^*)^{iE-1/2},
\ee
for $E\in\mathbb{R}$.
Hence the energy spectrum for a given $\kvec$, rather than being discrete as for the BL-sector, is the whole real line. 
In this representation we have the following two-point functions
\begin{align}
\langle 0|d_{\textbf{k}}b_{\textbf{k}'}|0\rangle&=-\langle 0|b_{\textbf{k}'}d_{\textbf{k}}|0\rangle=\frac{i}{2}\delta(\textbf{k}-\textbf{k}'),\\
\langle 0|d^\dagger_{\textbf{k}}b^\dagger_{\textbf{k}'}|0\rangle&=-\langle 0|b^\dagger_{\textbf{k}'}d^\dagger_{\textbf{k}}|0\rangle=\frac{i}{2}\delta(\textbf{k}-\textbf{k}').
\end{align}

To compute the Wightman function and the Feynman propagator we first rewrite our general
solution (\ref{soln4d}) as
\begin{align}\label{first}
\phi(x)& =-i\int d^4 k \delta^+ G(k)\left[a(k) e^{ik\cdot x}-a(k)^\dagger e^{-ik\cdot x}\right]-2\pi i\int \dk\left[-\frac{i}{8\pi^2}d_{k} e^{i\kappa\cdot x}-\frac{1}{\beta\Omega}b_{k}e^{-i\kappa\cdot x}\right]\\ \nonumber
& -2\pi i \int\dk\left[\frac{1}{{\beta^*\Omega^*}}b^\dagger_{k}e^{i\kappa^*\cdot x}-\frac{i}{8\pi^2}d^\dagger_{k}e^{-i\kappa^*\cdot x}\right]=:\varphi_{\text{BL}}(x)+\varphi(x)+\overline{\varphi}(x).
\end{align}
Since the BL and complex mass sectors do not mix (and neither
do $\varphi$ and $\overline{\varphi}$ within the complex-mass sector), 
we can look at their corresponding two-point functions separately. These are
\begin{align}
\langle 0 |\varphi_{\text{BL}}(x)\varphi_{\text{BL}}(y)\vac &=\fr{1}{4\pi}\int d^4k \delta^+G(k) e^{ik\cdot(x-y)}
=\fr{i}{4\pi}\int_{\Gamma_+} d^4k f^{-1}(k^2) e^{ik\cdot(x-y)},\\
\langle 0|\varphi(x)\varphi(y)|0\rangle & =\frac{1}{4}\int\dk\frac{1}{\beta\Omega}
\left(e^{i\kappa\cdot (x-y)}-e^{-i\kappa\cdot (x-y)}\right)
= \frac{i}{2}\int_{\Gamma_{1}} d^4 k\frac{\sin\left(k\cdot (x-y)\right)}{f(k^2)},\\
\langle 0|\overline{\varphi}(x)\;\overline{\varphi}(y)|0\rangle & =\frac{1}{4}\int\dk\frac{1}{\beta^*\Omega^*}\left(e^{i\kappa^*\cdot (x-y)}
-e^{-i\kappa^*\cdot (x-y)}\right)
= \frac{i}{2}\int_{\Gamma_{2}} d^4 k\frac{\sin\left(k\cdot (x-y)\right)}{f(k^2)}.
\end{align}
Hence the full Wightman function is given by
\begin{align}
\label{Wf}
\langle 0 |\phi(x)\phi(y)\vac =&\fr{1}{4\pi}\int d^4k \delta^+G(k) e^{ik\cdot(x-y)}\nn
&-\fr{i}{2}\int d^3\kvec e^{i\kvec\cdot(\bold{x}-\bold{y})}\left(\fr{1}{\beta\Omega}\sin(\Omega t)
-\fr{1}{\beta^*\Omega^*}\sin(\Omega^* t)\right),
\end{align}
and the Feynman propagator
\be
\langle 0 |T\{\phi(x)\phi(y)\}\vac = 
\frac{i}{4\pi}\int_{\Gamma_F} d^4 k \frac{e^{ik\cdot (x-y)}}{f(k^2)} + 
G_{W}(x-y) + G^*_{W}(x-y),
\label{feynman4d}
\ee
where 
\begin{align}
G_{W}(x) &=-\frac{i}{2}\text{sgn}(t)\int\dk \frac{1}{\beta\Omega}e^{-i\textbf{k}\cdot \textbf{x}}\sin\left(\Omega t\right),\\
G_W^*(x)&
=\frac{i}{2}\text{sgn}(t)\int\dk \frac{1}{\beta^*\Omega^*}e^{i\textbf{k}\cdot \textbf{x}}\sin\left(\Omega^* t\right),
\end{align}
are known as the Wheeler propagators for $\varphi$ and $\overline{\varphi}$
respectively,
and $\Gamma_F = \theta(x^0-y^0)\Gamma_+ - \theta(y^0-x^0)\Gamma_-$ is the Feynman 
contour defined in the previous section. 
The Wheeler propagator  
lacks the on-shell singular contribution of free field excitations (see Section 2 of \cite{Oxman:1992cm}).
\footnote{Effectively the Wheeler propagator is a sum of $1/f(k^2)$ with delta functions $\delta(k^0\pm\Omega^{(*)})$
such that the singular behaviour of $1/f$ at its complex roots is cancelled by the delta functions.}
Physically this means that these modes do not propagate asymptotically. 
\footnote{Wheeler and Feynman originally introduced this propagator in an attempt to provide relativistic
action-at-a-distance interpretation
of electrodynamics: the Wheeler-Feynman absorber theory \cite{RevModPhys.17.157}. 
One of the upshots of this interpretation was that it used both retarded and advanced propagators and
was therefore explicitly time-reversal invariant. 
In their theory charged particles
act as both emitters (via retarded solutions) and absorbers (via advanced solutions) of radiation,
hence the use of the Wheeler propagator. Nonetheless they were able to argue that,
provided there exist a sufficiently large number of charged particles in the universe absorbing radiation emitted by any one particle,
the overall field propagator is retarded, thus recovering causality. The physical picture of having a large number of (absorbing)
charges can be simply captured by imposing that the field vanish at infinity.
In our theory, the asymptotic absence of these modes arises by construction, i.e. by the way we decided
to quantise the complex mass sector. Whether this can be given sensible physical interpretation, other
than the fact that it ensures that these instabilities are not present asymptotically, remains an open issue whose
solution probably lies in the study of the interacting theory.
}

\subsection{Renormalisation
\label{ren}}

As was originally pointed out by Sorkin \cite{Sorkin:2007qi} the above propagator can be used  
to define a Lorentz invariant regularisation tool for QFT.
Indeed, as was fleshed out by Aslanbeigi {\it et al.} \cite{Aslanbeigi:2014tg},
the position space propagator $G(x,y)$ contains a $\delta$-function type singularity in the coincidence limit $y\rightarrow x$,
due to the constant term appearing in the UV expansion of $f(z)$, see Equation (3.19) in \cite{Aslanbeigi:2014tg}.
Subtracting this constant in momentum space leads to a regularised propagator, call it $G_{reg}(z)$, for which all loop integrals 
are finite. In the interacting theory one would therefore have to replace $f^{-1}$ with $G_{reg}$ in order to obtain 
a UV finite theory.

\section{Summary and Outlook}
\label{summary}

We have canonically quantised free massless scalar fields,
satisfying nonlocal equations of motion, in 2 and 4 
dimensions.
In both dimensions
we find a continuum
of massive modes arising from the cut discontinuity along $k^2\le0$ of the 
momentum space Green function, $f^{-1}(k^2)$, 
similar to what was found in previously studied free, non-local, massless QFTs \cite{Barci:1996kq}. 
This feature will be present in any dimension, since 
the existence of the branch cut is a generic feature of the Laplace transform
of retarded Lorentz invariant functions \cite{dominguez1979laplace}. 
In spite of the existence of these massive modes we showed that, in accordance with the 
analysis of Barnaby and Kamran \cite{Barnaby:2008ud}, only massless states appear
in the asymptotic state space of the quantum theory. 
In 4 dimensions, where the solution space to the non-local equations of motion is
augmented by the conjugate pair of complex mass modes, we found that by
constructing the quantum theory appropriately the states associated to these modes
are propagated via the Wheeler propagator, ensuring that 
they do not appear asymptotically and thus removing possible instabilities in the theory.
\footnote{It is interesting to note that Barnaby and Kamran \cite{Barnaby:2008ud}
suggest an alternative solution to the instabilities problem (
i.e.~the existence of complex mass solutions to (\ref{eom})), consisting in specifying a
contour in the complex $k^0$-plane which does not enclose the unstable modes. Their claim
being that a choice of contour is an integral part in the construction of a nonlocal QFT. If our
analysis of the unstable modes continues to stand strong once interactions are introduced, then it might not be
necessary to exclude such modes by hand.}

The above results lead to a wide range of interesting questions that in our mind require
further exploration. First, it would be interesting to find out what role the continuum of massive
modes play in the interacting theory, and whether the improved UV behaviour of the 4$d$ theory
(after renormalisation, see Section \ref{ren}) is due to its presence. 
To this end note that in 2 dimensions the discontinuity function 
$\dg$ is positive definite, ensuring that all massive modes in its support have positive energy, 
while in 4 dimensions it is not. It is natural therefore to ask whether the improved UV behaviour in 4
dimensions is related to this fact, or whether the existence of such 
negative energy states end up spoiling the theory irrevocably, e.g. by breaking unitarity.

Secondly, it remains to be shown that the complex mass modes present in the 4$d$ theory 
do not spoil unitarity themselves.
Similar work in this direction was done by Bollini and Oxman~\cite{Oxman:1993cm}, who 
showed 
that a higher order theory containing a conjugate pair of complex mass solutions is unitary. Whether their
analysis will continue to hold in the interacting theory, and whether it is applicable to our case remains to be 
investigated.

Thirdly it would be interesting to compare the above results with those of \cite{Johnston:2009fr} in which
the causet Feynman propagator is constructed. 
This could shed light on the Sweety-Salty duality defined in Section 3.2 of \cite{Johnston:2014sf}.
Doing this will require one to (at least) evaluate
the expressions for the Feynman propagator numerically. We intend to explore this in future work.

Finally we would like to speculate about the possible phenomenological consequences of these nonlocal
field theories. 
Recall from Section \ref{huygens} that one of the features of the 4$d$ theory is that 
it fails to satisfy Huygens' principle, i.e. there is a ``leakage"  inside the light cone of
the field emitted from a delta function source.
This could prove 
fruitful in testing the theory, since one can envisage performing high precision tests of  radiation emitted
by very localised sources to check if such afterglow is present.
\footnote{Clearly our analysis strictly only applies to scalar fields, however one can imagine
that similar features will carry over to higher spin fields. } 
A key feature of this phenomenon, due to its intrinsic
Lorentz invariance, is that it will be frequency independent.
Another interesting phenomenological avenue is to study the non-relativistic limit of (\ref{fmass}). It is well known
that the Schr\"odinger equation is the non-relativistic limit of the Klein-Gordon equation. Therefore, the non-relativistic
limit of (\ref{fmass}) will give rise to a non-local generalisation of the Schr\"odinger equation whose non-locality is again
parametrised by $\rho$ and is such that the usual Schr\"odinger equation is recovered in the limit $\rho\rightarrow\infty$ 
\cite{Belenchia}.

\acknowledgments

The authors would like to thank Luis Oxman for helpful comments during early stages of the project and 
Michel Buck for comments on earlier drafts. 
Special thanks also go to Leonardo Modesto, Domenico Giulini, Bruno Lima de Souza, Sebastiano Sonego
and Marco Letizia.
This publication was made possible through the support of a grant from the John Templeton 
Foundation. The opinions expressed in this publication are those of the authors and do not 
necessarily reflect the views of the John Templeton Foundation.


\appendix

\section{Degree of Singularity of Complex Mass Poles
\label{appC}}

In this appendix we provide evidence that the complex mass poles appearing in the 4$d$ propagator are of order 1. 

\begin{figure}[H]
\centering
\includegraphics[scale=0.6]{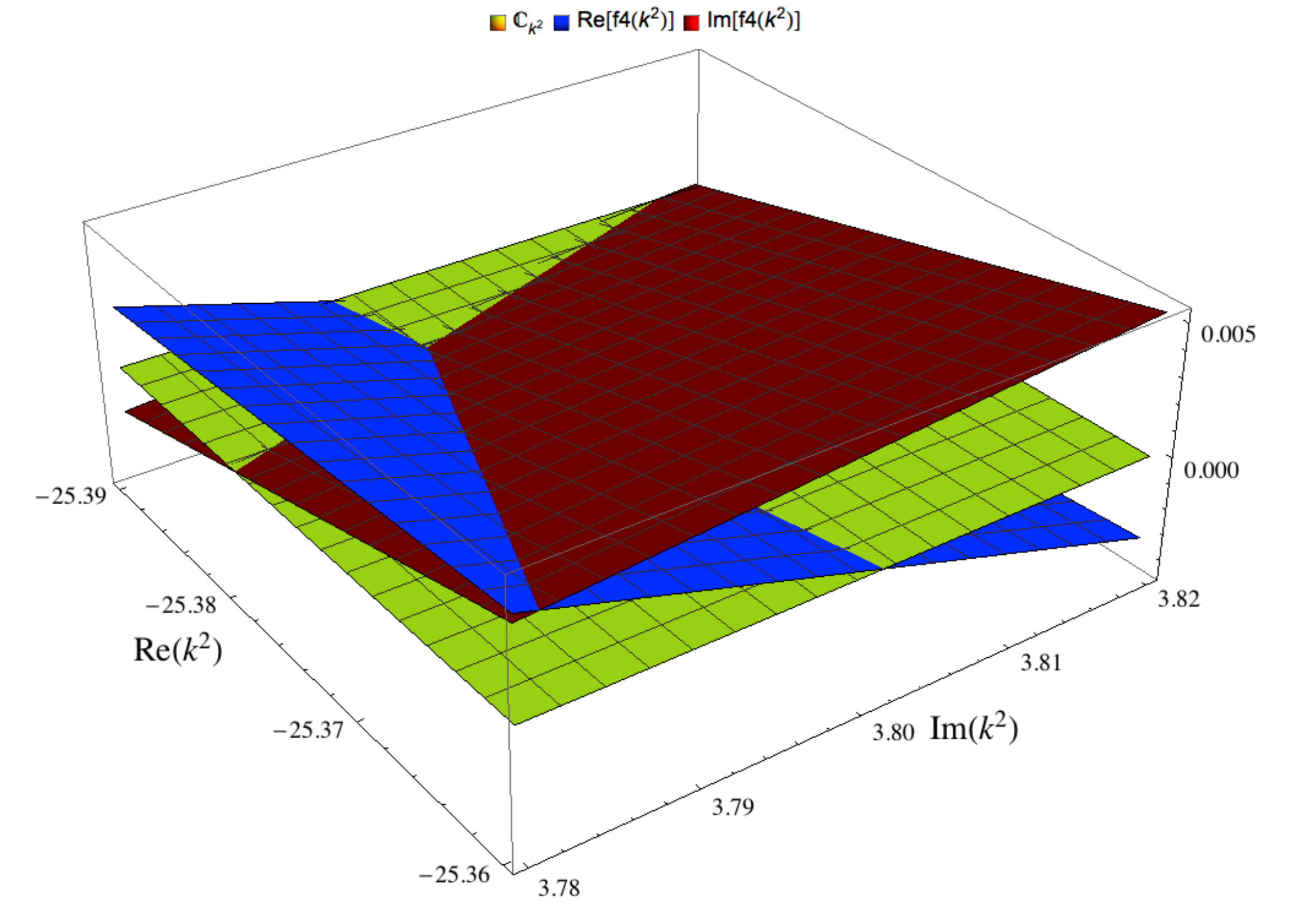}
\caption{Real and imaginary part together. The two planes intersect with the complex-plane in what seems like a single 
point representing one of the complex roots of $f^{(4)}(z)$.}\label{ReIm}
\end{figure}

The existence of these poles was established numerically using Cauchy's argument principle, as such we only know
their location in the complex $k^2$ plane approximately. In order to establish their nature 
we probe the behaviour of $f(z)$ in the vicinity of these zeros numerically, and provide evidence  
that $f(z)$ is linear in $z$ in these neighbourhoods (see Figure\ref{ReIm}). A linear behaviour in $z$ corresponds to the inverse having simple 
poles.

\section{Series Form of the Non-Local Operator
\label{appA}}
In this appendix we show how to rewrite the Laplace transform of the 2$d$ and 4$d$ 
non-local d'Alembert operator in series form. This allows one to easily determine the IR corrections to the exact continuum
d'Alembertian.\par

\subsection{4 Dimensions}

Following the notation in \cite{Aslanbeigi:2014tg} we begin by rewriting  Equation (\ref{dalemb4d})
\be
\label{4dSform}
\tilde{f}^{(4)}(Z):=-\fr{4}{\sqrt{6}}+\fr{16}{\sqrt{6}}\pi Z^{-1/2}\int_0^{\infty} ds \,s^{2} e^{- C_4 s^{4}}K_{1}\left(Z^{1/2}s\right)\left(1-9 C_4 s^{4}+8 C_{4}^{2} s^8-\fr{4}{3}C_{4}^3 s^{12}\right),
\ee
where $f^{4}(k^2)=\sqrt{\rho}\tilde{f}^{(4)}(Z)$, $Z=k^{2}/\sqrt{\rho}$ and $C_{4}=\pi/24$. The IR condition then takes the form $Z\ll 1$.
From the power series expansion of $K_{1}$ (see Equation 10.31.1 of \cite{Olver:2010:NHMF}) one has
\begin{align}
(Z^{1/2}s)^{-1}K_{1}(Z^{1/2}s)&=(Z s^{2})^{-1}+\frac{1}{2}\ln\left(\frac{1}{2}Z^{1/2}s\right)\sum_{k=0}^{\infty}\frac{\left(\frac{1}{4}Z s^{2}\right)^{k}}{k! \Gamma(k+2)}\\ \nonumber
& -\frac{1}{4}\sum_{k=0}^{\infty}\left(\psi(k+1)+\psi(k+2)\right)\frac{\left(\frac{1}{4}Z s^{2}\right)^{k}}{k! (k+1)!}.
\end{align}  
As shown in Appendix A of \cite{Aslanbeigi:2014tg}, the first two terms of this series ($k=0,1$), 
when substituted back into Equation (\ref{4dSform}), give $-Z$
. The procedure can be continued for generic values of 
$k$ in order to obtain the series form of the entire Laplace transform of the non-local d'Alembertian. 
The series can be compactly written as 
\be
\tilde{f}^{(4)}(Z)=-Z+\sum_{k=2}^{\infty}c_k(Z)Z^{k},
\ee  
where the $k$th term of the series is given by,
\begin{align}
c_k(Z) &= -\frac{1}{\Gamma(k+2)^{2}}2^{-\frac{k+1}{2}}3^{-\frac{k+1}{2}}\pi^{-\frac{k}{2}}\left(16(k+1)\Gamma(\frac{k}{2}+3)\right.\nn
&\left.+k^{3}\Gamma(\frac{k}{2})\left(\ln(2\pi/3)-k^{2}\ln(2\pi/3)-2k-k\ln(2\pi/3)+2(k^{2}+k-1)\ln(Z)\right)
\right.\nn 
&\left.
+\Gamma(1+\frac{k}{2})(-36-56k-16k^2 +8k^3 +k\ln(\frac{4\pi^{2}}{9})-4k\ln(Z)
\right.\nn
&\left.+2k(k-1)(k+1)^2 \psi(1+\frac{k}{2})-8k(k-1)(k+1)^2 \psi(1+k))\right).
\end{align}
Note that $c_k$ can be written as $c_k=a_k+b_k\ln(Z)$ for constants $a_k$ and $b_k$.

\subsection{2 Dimensions}
In 2$d$ we use the power series expansion of $K_{0}$ given by
\be
K_{0}\left(z^{1/2}\right)=-\ln\left(\frac{1}{2}z^{1/2}s\right)\sum_{k=0}^{\infty}\frac{\left(\frac{1}{4}z s^2\right)^k}{k!\Gamma(k+1)}
+\frac{1}{2}\sum_{k=0}^{\infty}2\psi(k+1)\frac{\left(\frac{1}{4}z s^2\right)^k}{(k!)^2},
\ee
to solve for 
\be
\tilde{f}^{(2)}(Z)= \frac{f^{2}(k^{2})}{\rho}=-2+2\int_{0}^{\infty} ds \,s\, e^{-s^2 /2}\left(4-4s^2 +\frac{1}{2}s^4 \right) K_{0}\left(Z^{1/2}s\right),
\label{f2d}
\ee
order by order.
The terms corresponding to $k=0,1$ in the expansion of the modified Bessel function combine 
together to give the IR behaviour of (\ref{f2d}): 
\be
\tilde{f}^{(2)}(Z)=-Z+\dots.
\ee
The higher order terms give rise to a power series in which the $k$th term has the form:
\be
a_{k} Z^{k}+b_{k}\ln(Z)Z^{k}:=\frac{2^{1-k}Z^{k}(1-2k+k(k-1) \ln(2)-k(k-1)\ln(Z)+k(k-1)\psi(k+1))}{\Gamma(k+1)}.
\ee 
Hence\be
\tilde{f}^{(2)}(Z)=-Z+\sum_{k=2}^{\infty} \left(a_{k}+b_{k}\ln(Z)\right) Z^{k}.
\ee
Equations 
(\ref{2dEx}) and (\ref{4dEx}) can then be easily obtained by truncating these series at $k=2$.

\section{Asymptotic States
\label{appB}}

Here we show that asymptotic states of the BL sector only contain massless
states. We follow the analysis of Barci and Oxman \cite{Barci:1997xy}.

The fields $\phi(x)$, solutions to the nonlocal equations of motion (\ref{eom}), 
are examples of {\it generalised free fields}, introduced in \cite{Greenberg:1961if}.
\footnote{In 4 dimensions the definitions don't fully match since the generalised free fields of
Greenberg have to have positive weight $\delta^+G$. This condition is not essential
for this part of the analysis, but its relaxation will need to be dealt with carefully 
once interactions are introduced.}
These are said to be associated with asymptotic modes of mass $m$ if the asymptotic
in/out fields
\begin{equation}
\phi_{m^2}^{\pm}(x):= \lim_{\tau\rightarrow\pm\infty}\int_{y^{0}=\tau}\dy \left(\phi(y)\overleftrightarrow{\Dy}\Dm(x-y)\right)
\end{equation}
satisfy the free field commutation relations
\begin{equation}
\left[\phi_{m^2}^{\pm}(x),\phi_{m^2}^{\pm}(y)\right]=i\Dm(x-y).
\label{comm}
\end{equation}

Recall that the general solution to the nonlocal equations of motion (\ref{eom})
in the BL sector is given by
\begin{align}
\phi(x)& =\int d^d k\, \delta^{+}G(k)\left(a(k) e^{i k\cdot x}-a(k)^\dagger e^{-i k\cdot x}\right).
\end{align}
Then the Pauli-Jordan function is
\begin{equation}
 i\Delta_{NL}(x-y)\equiv\left[\phi(x),\phi(y)\right]=\frac{i}{4\pi}\int d\mu^2 \delta^{+}G(\mu^2)\Dmu(x-y),
\end{equation}
where $\Dmu(x-y)$ is the Pauli-Jordan function for a free field of mass $\mu$ satisfying
the KG equation.
Now,
\begin{align}
&\left[\phi_{m^2}^{\pm}(x),\phi_{m^2}^{\pm}(y)\right] =\lim_{\tau,\tau'\rightarrow\pm\infty}\int_{z^{0}=\tau}\!\!\!\!\!\!\dz\int_{z'^{0}=\tau}\!\!\!\!\!\!\dzp
\Big\{\left[\phi(z),\phi(z')\right]\Dz\Dm(x-z)\Dzp\Dm(y-z')\nn
&-\left[\phi(z),\pi(z')\right]\Dz\Dm(x-z)\Dm(y-z')
-\left[\pi(z),\phi(z')\right]\Dm(x-z)\Dzp\Dm(y-z')\nn
& +\left[\pi(z),\pi(z')\right]\Dm(x-z)\Dm(y-z')\Big\}.
\end{align}
The four terms in the integrand are
\begin{align}
(1)&=\frac{i}{4\pi}\int d\mu^2 \delta^{+}G(\mu^2)\Dmu(z-z')\Dz\Dm(x-z)\Dzp\Dm(y-z'),\\
(2)&=-\frac{i}{4\pi}\int d\mu^2 \delta^{+}G(\mu^2)\Dzp\Dmu(z-z')\Dz\Dm(x-z)\Dm(y-z'),\\
(3)&=-\frac{i}{4\pi}\int d\mu^2 \delta^{+}G(\mu^2)\Dz\Dmu(z-z')\Dm(x-z)\Dzp\Dm(y-z'),\\
(4)&=\frac{i}{4\pi}\int d\mu^2 \delta^{+}G(\mu^2)\Dz\Dzp\Dmu(z-z')\Dm(x-z)\Dm(y-z').
\end{align}
Integrating the sum of the first two terms and taking the limit gives
\begin{align}
\frac{i}{4\pi}\lim_{\tau\rightarrow\pm\infty}\int_{z^{0}=\tau}\dz\Dz\Dm(x-z)\int d\mu^2 \delta^{+}G(\mu^2)\begin{cases} \Dmu(z-y), & \mbox{if }m^{2}=\mu^{2} \\ 0, & \mbox{otherwise }
\end{cases}
\end{align}
Doing the same for the sum of the last two terms
\begin{align}
\frac{i}{4\pi}\lim_{\tau\rightarrow\pm\infty}\int_{z^{0}=\tau}\dz\Dm(x-z)\Dz\int d\mu^2 \delta^{+}G(\mu^2)\begin{cases}-\Dmu(z-y), & \mbox{if }m^{2}=\mu^{2} \\ 0, & \mbox{otherwise },
\end{cases}
\end{align}
where we used the fact $\Dmu(y-z)=-\Dmu(z-y)$.
Putting these together we find
\begin{align}
-\frac{i}{4\pi}\lim_{\tau\rightarrow\pm\infty}\int_{z^{0}=\tau}\dz\int d\mu^2 \delta^{+}G(\mu^2)\Dm(x-z)\overleftrightarrow{\Dz}
\begin{cases} \Dmu(z-y), & \mbox{if }\mu^2=m^2 \\ 0, & \mbox{otherwise }.
\end{cases}
\label{asymp}
\end{align}
Since $\mu^2=m^2$ is a set of measure zero, Equation (\ref{asymp}) is non zero only
at delta-function type singularities of $\dg$, i.e. for $\mu^2=0$. 
Using the following identity 
\begin{equation}
\lim_{\tau\rightarrow\pm\infty}\int_{y^{0}=\tau}\dy 
\left(\Dmz(x-y)\overleftrightarrow{\Dy}\Dm(z-y)\right)=\begin{cases} 
\Dmz(x-z), & \mbox{if }m^{2}=m_{0}^{2} \\ 0, & \mbox{otherwise },
\end{cases}
\label{identity}
\end{equation} 
we find that the asymptotic in/out fields satisfy
massless free field commutation relations
\begin{equation}
\left[\phi_{0}^{\pm}(x),\phi_{0}^{\pm}(y)\right]=i\Delta_0(x-y).
\end{equation}

\bibliographystyle{apsrev}
\bibliography{refs}

\end{document}